\documentclass[
reprint,
superscriptaddress,
showpacs,preprintnumbers,
 amsmath,amssymb,
 aps,
 prc,
]{revtex4-1}
\usepackage{subcaption}
\usepackage{graphicx}
\usepackage{dcolumn}
\usepackage{bm}
\usepackage{ upgreek }
\usepackage{units}
\usepackage[utf8]{inputenc}
\usepackage[T1]{fontenc}
\usepackage[colorlinks=true, allcolors=blue]{hyperref}

\begin{document}

\preprint{APS/123-QED}

\title{Precision mass measurements on neutron-rich rare-earth isotopes at JYFLTRAP - reduced neutron pairing and implications for the $r$-process calculations}

\author{M.~Vilén}
\email{markus.k.vilen@student.jyu.fi}
\affiliation{University of Jyv{\"a}skyl{\"a}, P.O. Box 35, FI-40014 University of Jyv{\"a}skyl{\"a}, Finland}
\author{J.M.~Kelly}
\email{jkelly27@nd.edu}
\affiliation{University of Notre Dame, Notre Dame, Indiana 46556, USA}
\author{A.~Kankainen}
\affiliation{University of Jyv{\"a}skyl{\"a}, P.O. Box 35, FI-40014 University of Jyv{\"a}skyl{\"a}, Finland}
\author{M.~Brodeur}
\affiliation{University of Notre Dame, Notre Dame, Indiana 46556, USA}
\author{A.~Aprahamian}
\affiliation{University of Notre Dame, Notre Dame, Indiana 46556, USA}
\author{L.~Canete}
\affiliation{University of Jyv{\"a}skyl{\"a}, P.O. Box 35, FI-40014 University of Jyv{\"a}skyl{\"a}, Finland}
\author{T.~Eronen}
\affiliation{University of Jyv{\"a}skyl{\"a}, P.O. Box 35, FI-40014 University of Jyv{\"a}skyl{\"a}, Finland}
\author{A.~Jokinen}
\affiliation{University of Jyv{\"a}skyl{\"a}, P.O. Box 35, FI-40014 University of Jyv{\"a}skyl{\"a}, Finland}
\author{T.~Kuta}
\affiliation{University of Notre Dame, Notre Dame, Indiana 46556, USA}
\author{I.D.~Moore}
\affiliation{University of Jyv{\"a}skyl{\"a}, P.O. Box 35, FI-40014 University of Jyv{\"a}skyl{\"a}, Finland}
\author{M.R.~Mumpower}
\affiliation{University of Notre Dame, Notre Dame, Indiana 46556, USA}
\affiliation{Theory Division, Los Alamos National Lab, Los Alamos, New Mexico 87544, USA}
\author{D.A.~Nesterenko}
\affiliation{University of Jyv{\"a}skyl{\"a}, P.O. Box 35, FI-40014 University of Jyv{\"a}skyl{\"a}, Finland}
\author{H.~Penttil{\"a}}
\affiliation{University of Jyv{\"a}skyl{\"a}, P.O. Box 35, FI-40014 University of Jyv{\"a}skyl{\"a}, Finland}
\author{I.~Pohjalainen}
\affiliation{University of Jyv{\"a}skyl{\"a}, P.O. Box 35, FI-40014 University of Jyv{\"a}skyl{\"a}, Finland}
\author{W.S.~Porter}
\affiliation{University of Notre Dame, Notre Dame, Indiana 46556, USA}
\author{S.~Rinta-Antila}
\affiliation{University of Jyv{\"a}skyl{\"a}, P.O. Box 35, FI-40014 University of Jyv{\"a}skyl{\"a}, Finland}
\author{R. Surman}
\affiliation{University of Notre Dame, Notre Dame, Indiana 46556, USA}
\author{A.~Voss}
\affiliation{University of Jyv{\"a}skyl{\"a}, P.O. Box 35, FI-40014 University of Jyv{\"a}skyl{\"a}, Finland}
\author{J.~\"Ayst\"o}
\affiliation{University of Jyv{\"a}skyl{\"a}, P.O. Box 35, FI-40014 University of Jyv{\"a}skyl{\"a}, Finland}

\date{\today}

\begin{abstract}
The rare-earth peak in the $r$-process abundance pattern depends sensitively on both the astrophysical conditions and subtle changes in nuclear structure in the region. This work takes an important step elucidating the nuclear structure and reducing the uncertainties in $r$-process calculations via precise atomic mass measurements at the JYFLTRAP double Penning trap. $^{158}$Nd, $^{160}$Pm, $^{162}$Sm, and $^{164-166}$Gd have been measured for the first time and the precisions for $^{156}$Nd, $^{158}$Pm, $^{162,163}$Eu, $^{163}$Gd, and $^{164}$Tb have been improved considerably. Nuclear structure has been probed via two-neutron separation energies $S_{2n}$ and neutron pairing energy metrics $D_n$. The data do not support the existence of a subshell closure at $N=100$. Neutron pairing has been found to be weaker than predicted by theoretical mass models. The impact on the calculated $r$-process abundances has been studied. Substantial changes resulting in a smoother abundance distribution and a better agreement with the solar $r$-process abundances are observed.   
\end{abstract}

\maketitle

The astrophysical rapid neutron capture process ($r$-process) \cite{Burbidge1957,Cameron1957,Arnould2007} is responsible for the production of around half of the elements heavier than iron. The $r$-process and its astrophysical site has driven research not only in nuclear astrophysics, but in multiple fields, including nuclear structure \cite{Pfeiffer2001,Sorlin2008} and theory \cite{Erler2012,Martin2016}, accelerator mass spectrometry \cite{Wallner2015} and observational astronomy \cite{Ji2016, Roederer16}. Various astrophysical sites have been proposed over the years, e.g. neutrino-driven winds from the remnants of core-collapse supernovae \cite{Arnould2007,Arcones2013}, magnetohydrodynamic supernovae \cite{Mosta2015}, and neutron star mergers \cite{Lattimer1974,Argast2004,Freiburghaus1999,Goriely2011,Goriely2013b,Thielemann2017}. The recent, seminal multi-messenger observations of a neutron star merger \cite{Coulter2017, Abbott2017b}, namely the gravitational waves from GW170817 \cite{Abbott2017} followed by a kilonova (AT 2017 gfo) powered by the radioactive decay of $r$-process nuclei synthesized in the ejecta \cite{Arcavi2017,Kasliwal2017}, provide direct evidence that the $r$-process takes place in neutron star mergers. For the first time, this allows the testing of $r$-process abundance models using an unpolluted sample \cite{Kasen2017}. Hence, there is now a strong impetus to have accurate nuclear physics inputs to ensure the reliability of the abundance calculations. With their high opacity, lanthanides play a central role in the diagnostics of heavy $r$-process ejecta from such mergers \cite{Tanvir2017,Cowperthwaite2017}. In this Letter, we present results for nuclear binding energies that affect the calculated $r$-process abundances of lanthanides in the rare-earth region. 

Because the $r$-process path traverses uncharted and largely inaccessible regions of the chart of nuclides, there is a scarcity of experimental information with which to constrain the astrophysical calculations. Detailed $r$-process sensitivity studies performed in recent years \cite{Aprahamian2014,Brett2012,Mumpower2016,Mumpower2017,mumpower2015prc035807,mumpower2015jphysg034027} have shown that among the various quantities entering into their calculations, e.g. neutron-capture and photodisintegration rates, beta-decay half-lives, and beta-delayed neutron emission and fission probabilities, it is the quantities most strongly derivative of nuclear mass, namely binding energies, that proved to be the most sensitive \cite{Mumpower2016}. However, the masses of the most relevant $r$-process nuclei have never been measured, leaving nuclear abundance calculations to rely on theoretical mass models such as FRDM12 \cite{Moller2016}, HFB-24 \cite{Goriely2013}, Duflo-Zuker \cite{duflo1995} or Skyrme energy-density functionals \cite{Martin2016} for these critical inputs. While the mass models agree closely with one another in regions with existing measurements, they diverge strongly in the absence of such empirical data, which has profound impacts on abundance peak formation simulations \cite{Mumpower2016}. 

The formation of the rare-earth abundance peak is very sensitive to nuclear structure in the neutron-rich rare-earth region. A confluence of nuclear deformation and $\beta$-decay properties peculiar to nuclei surrounding $A=165$ is understood to create a funneling effect that draws the nuclei towards the peak as neutron captures dwindle and existing radionuclides decay towards stability \cite{Surman1997,mumpower2012_RE}. Furthermore, fission recycling is believed to augment this process as the fragments of heavier, unstable nuclides beyond the third ($A\approx 195$) peak could cycle back into the rare-earth region \cite{beun2008,mumpower2012_RE,Goriely2013b}.  Fortunately, the rare-earth abundances are some of the most precisely known in the solar system and in metal-poor stars \cite{lodders2009}. 



The rare-earth region, located in the midshell bounded by $Z=50-82$ and $N=82-126$, incorporates several interesting nuclear structure features that can affect the $r$-process. A surge of research was triggered by the discovery of the onset of strong prolate deformation at $N=88-90$ in the 1950s \cite{Brix1952,Mottelson1955}. Proton-neutron interactions enhanced in nuclei with approximately equal numbers of valence protons and neutrons have been found to play a key role in the evolution of nuclear structure and collectivity in this region \cite{Casten1981,Bonatsos2013,Casten2006}. A local minimum in the $E(2^+)$ energies and a local maximum of moment of inertia have been observed for the Gd isotopes at $N=98$ via $\gamma$-ray spectroscopy at Gammasphere \cite{Jones2005}. Jones et al. \cite{Jones2005} found $^{164}$Gd ($N=100$) to be more rigid and to show less stretching than $^{162}$Gd, suggesting a possible change in structure. Recently, $\gamma$-ray spectroscopy on $^{164}$Sm and $^{166}$Gd with EURICA at RIBF revealed an increase in the $E(2^+)$ and $E(4^+)$ energies at $N=100$ in comparison with the $N=98$ cases for Gd and Sm isotopes, supporting an implied sub-shell closure at $N=100$ proffered by the Hartree-Fock calculations of \cite{Ghorui2012}. Interestingly, recent half-life measurements performed at RIKEN \cite{wu2017} did not find any supporting evidence for the $N=100$ subshell closure. Additionally the systematics of the new $K$ isomers found in the neutron-rich $N=100$ isotones $^{162}$Sm, $^{163}$Eu, and $^{164}$Gd at RIKEN could be explained without the predicted $N=100$ shell gap \cite{Yokoyama2017}.
      
Although information on beta-decay half-lives \cite{wu2017} and level structures \cite{Jones2005,Patel2014} of rare-earth nuclei has increased substantially in recent years, nuclear binding energies - i.e. masses - have not been pursued so intensively. The Canadian Penning Trap (CPT) has explored some rare-earth nuclei in the past \cite{vanschelt2012}, and some $Q_{\beta}$ measurements have been performed using a total absorption Clover detector \cite{hayashi2014}. In this Letter we present the first mass measurements of several rare-earth nuclei close to $N=100$ of significance for the astrophysical $r$-process, while providing further information on the nuclear structure which is of direct relevance for the $r$-process.

\begin{figure}
\centering
\includegraphics[trim = 65mm 100mm 75mm 115mm, clip, width=0.48\textwidth]{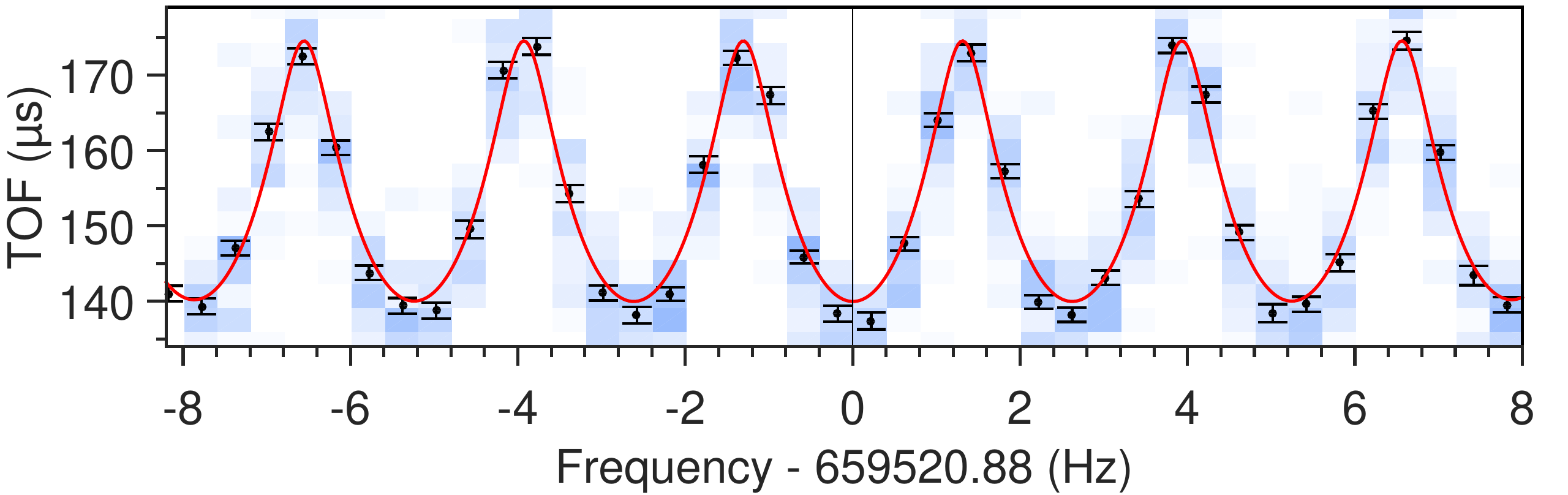}
\caption{(Color online) Time-of-flight spectrum for $^{163}\mathrm{Eu}^+$. Background shading indicates the total number of ions, where darker shading indicates more ions.}
\label{figure_tof}
\end{figure}

The studied neutron-rich rare-earth nuclei were produced at the Ion Guide Isotope Separator On-Line (IGISOL) facility \cite{igisol}, employing a $\unit[25]{MeV}$, $10-15~\mu A$ proton beam impinging on a $\unit[15]{mg/cm^{2}}$-thick natural uranium target. The fission fragments were thermalized in helium buffer gas and extracted from the gas cell with a typical charge state of $q =+e$ by a radio-frequency sextupole ion guide \cite{spig}. Subsequently, the ions were accelerated to $\unit[30]{keV}$ before mass-separation with a dipole magnet. The continuous beam was cooled and bunched in a radio-frequency quadrupole cooler-buncher (RFQ) \cite{rfq} prior to injection into the double Penning trap mass spectrometer, JYFLTRAP \cite{Eronen2012}. Isobarically pure ion samples were prepared in the purification trap via the mass-selective buffer gas cooling method \cite{Savard1991}. For $^{156}\mathrm{Nd}$, $^{158}\mathrm{Pm}$, $^{162}\mathrm{Sm}$, $^{162-163}\mathrm{Eu}$, $^{163-166}\mathrm{Gd}$ and $^{164}\mathrm{Tb}$, an additional cleaning phase employing dipolar Ramsey excitations \cite{r-cleaning} in the second trap was required. The mass measurements were performed by determining the cyclotron frequency, $\nu_{c} = qB/(2\pi m)$, for an ion with mass $m$ and charge $q$ in a magnetic field $B$ using the time-of-flight ion-cyclotron resonance method (TOF-ICR) \cite{Graff1980,Konig1995} (see Fig. \ref{figure_tof}). A 400-ms quadrupolar excitation scheme was applied for $^{158}\mathrm{Nd}$ and $^{160}\mathrm{Pm}$. To more accurately determine the frequency, separated oscillatory fields \cite{ramsey2,ramsey1} with excitation patterns of 25-350-25~ms and 25-750-25~ms (On-Off-On) were applied for $^{156}\mathrm{Nd}$, $^{158}\mathrm{Pm}$, $^{162}\mathrm{Sm}$, $^{162-163}\mathrm{Eu}$, $^{163-166}\mathrm{Gd}$, and $^{164}\mathrm{Tb}$.

\begin{table*}
\caption{\label{tab:table3} Frequency ratios ($r$) and mass-excess values ($ME$) determined in this work with JYFLTRAP compared with AME16 \cite{AME16}. All measurements were done with singly-charged ions. The reference masses, $^{136}\mathrm{Xe}$, $^{158}\mathrm{Gd}$, $^{163}\mathrm{Dy}$, and $^{171}\mathrm{Yb}$, were adopted from AME16, and \# signs indicate extrapolated values therein.}
\begin{ruledtabular}
\begin{tabular}{lllllll}
Isotope & Reference & $ME_{REF}\mathrm{(keV)}$ & $r=\nu_{c,ref}/\nu_{c}$ & $ME_{JYFL}\mathrm{(keV)}$ & $ME_{AME16}\mathrm{(keV)}$ & $\Delta ME_{JYFL - AME16}\mathrm{(keV)}$\\ 
\hline
$^{156}\mathrm{Nd}$ & $^{136}\mathrm{Xe}$ & -86429.159(7) & $1.147\ 366\ 924(19)$ & -60210(2) & -60470(200) & 260(200)  \\
$^{158}\mathrm{Nd}$ & $^{136}\mathrm{Xe}$ & -86429.159(7) & $1.162\ 132\ 772(290)$ & -53897(37) & -54060(200)\# & 160(200)\# \\
$^{158}\mathrm{Pm}$ & $^{158}\mathrm{Gd}$ & -70689.5(12) & $1.000\ 078\ 752(9)$ & -59104(2) & -59089(13) & -15(13) \\
$^{160}\mathrm{Pm}$ & $^{136}\mathrm{Xe}$ & -86429.159(7) & $1.176\ 857\ 014(130)$ & -52851(16) & -53000(200)\# & 149(201)\# \\
$^{162}\mathrm{Sm}$ & $^{136}\mathrm{Xe}$ & -86429.159(7) & $1.191\ 560\ 914(39)$ & -54381(5) & -54530(200)\# & 149(200)\#  \\
$^{162}\mathrm{Eu}$ & $^{136}\mathrm{Xe}$ & -86429.159(7) & $1.191\ 527\ 132(28)$ & -58658(4) & -58700(40) & 42(40) \\
$^{163}\mathrm{Eu}$ & $^{163}\mathrm{Dy}$ & -66381.2(8) & $1.000\ 065\ 633(23)$ & -56420(4) & -56480(70) & 60(70) \\
$^{163}\mathrm{Gd}$ & $^{163}\mathrm{Dy}$ & -66381.2(8) & $1.000\ 034\ 135(22)$ & -61200(4)\footnote{Assuming the measured state is the isomer at 137.8 keV \cite{hayashi2014}, the ground-state mass is $-61338(4)$~keV.} & -61314(8) & 114(9) \\
$^{164}\mathrm{Gd}$ & $^{171}\mathrm{Yb}$ & -59306.810(13) & $0.959\ 046\ 522(14)$ & -59694(3) & -59770(100)\# & 76(100)\# \\
$^{165}\mathrm{Gd}$ & $^{171}\mathrm{Yb}$ & -59306.810(13) & $1.058\ 489\ 243(23)$\footnote{Measured as $^{165}\mathrm{Gd}^{16}\mathrm{O}$.} & -56522(4) & -56450(120)\#	& -72(120)\# \\
$^{166}\mathrm{Gd}$ & $^{136}\mathrm{Xe}$ & -86429.159(7) & $1.220\ 992\ 828(29)$ & -54387(4) & -54530(200)\# & 143(200)\#\\
$^{164}\mathrm{Tb}$ & $^{171}\mathrm{Yb}$ & -59306.810(13) & $0.959\ 031\ 473(21)$ & -62090(4) & -62080(100) & -10(100) \\
\end{tabular}
\end{ruledtabular}
\label{results}
\end{table*}

The magnetic field strength was precisely determined by interleaving measurements of a well-known reference ion ($\nu_{c,ref}$) just before and after an ion of interest ($\nu_{c}$). The mass ratios and atomic masses were then calculated from the ratio of frequencies $r=\nu_{c,ref}/\nu_{c}$, which equals the ratio of the ion masses. Data analysis followed the procedure described in \cite{Weber2008,Eronen2012}. Temporal fluctuations of the $B$-field, $\delta_{B}(\nu_{ref})/\nu_{ref} = \Delta t\cdot\unit[8.18\cdot 10^{-12}]{/min}$ \cite{Berror}, where $\Delta t$ is the time between consecutive reference measurements, and a mass-dependent uncertainty $\delta_{m}(r)/r = \Delta m \cdot \unit[2.2(6) \cdot 10^{-10}]{/u}$, determined soon after the experiment, were taken into account.

The measured frequency ratios and the corresponding mass-excess values are presented in Table~\ref{results}. Six isotopes, namely $^{158}\mathrm{Nd}$, $^{160}\mathrm{Pm}$, $^{162}\mathrm{Sm}$, $^{164-166}\mathrm{Gd}$, were measured for the first time. The precision of the mass values has been improved considerably for all studied isotopes. The new values agree with the extrapolations of AME16 \cite{AME16}, which have generally overestimated the nuclear binding energies in this region by about 150~keV. 

Most of the previously known mass values were based on $\beta$-decay $Q$-value measurements, such as $^{156}\mathrm{Nd}$ \cite{Shibata2003}, $^{162,163}\mathrm{Eu}$ \cite{hayashi2014}, and $^{164}\mathrm{Tb}$ \cite{Gujrathi71}. Although the $Q_\beta$ values yield lower mass values than the present Penning trap measurement, only $^{156}\mathrm{Nd}$ \cite{Shibata2003} deviates by more than $1\sigma$ from this work. In fact, it has been suggested \cite{AME16_evaluation} that based on the trends on the mass surface, $^{156}\mathrm{Nd}$ might actually be $\unit[70]{keV}$ less bound. 

Two of the studied isotopes, $^{158}\mathrm{Pm}$ and $^{163}\mathrm{Gd}$, have been measured by CPT \cite{vanschelt2012}. While the results for $^{158}\mathrm{Pm}$ agree within $1\sigma$, they deviate considerably in the case of $^{163}\mathrm{Gd}$. Interestingly, a new long-lived ($T_{1/2}=\unit[23.5(10)]{s}$) isomeric state at $\unit[137.8]{keV}$ in $^{163}\mathrm{Gd}$ was recently discovered \cite{hayashi2014}. The unusually large discrepancy between this work and CPT \cite{vanschelt2012} could be understood if the proton-induced fission on $^{nat}\mathrm{U}$ at IGISOL had predominantly populated the isomeric state of $^{163}\mathrm{Gd}$. Assuming we measured the first isomeric state, our corrected mass-excess value differs from CPT by 24(9)~keV. If we use the 15 keV uncertainty quoted in \cite{vanschelt2012} rather than AME16, it results in an even better agreement.

\begin{figure}
\centering
\includegraphics[width=0.90\linewidth]{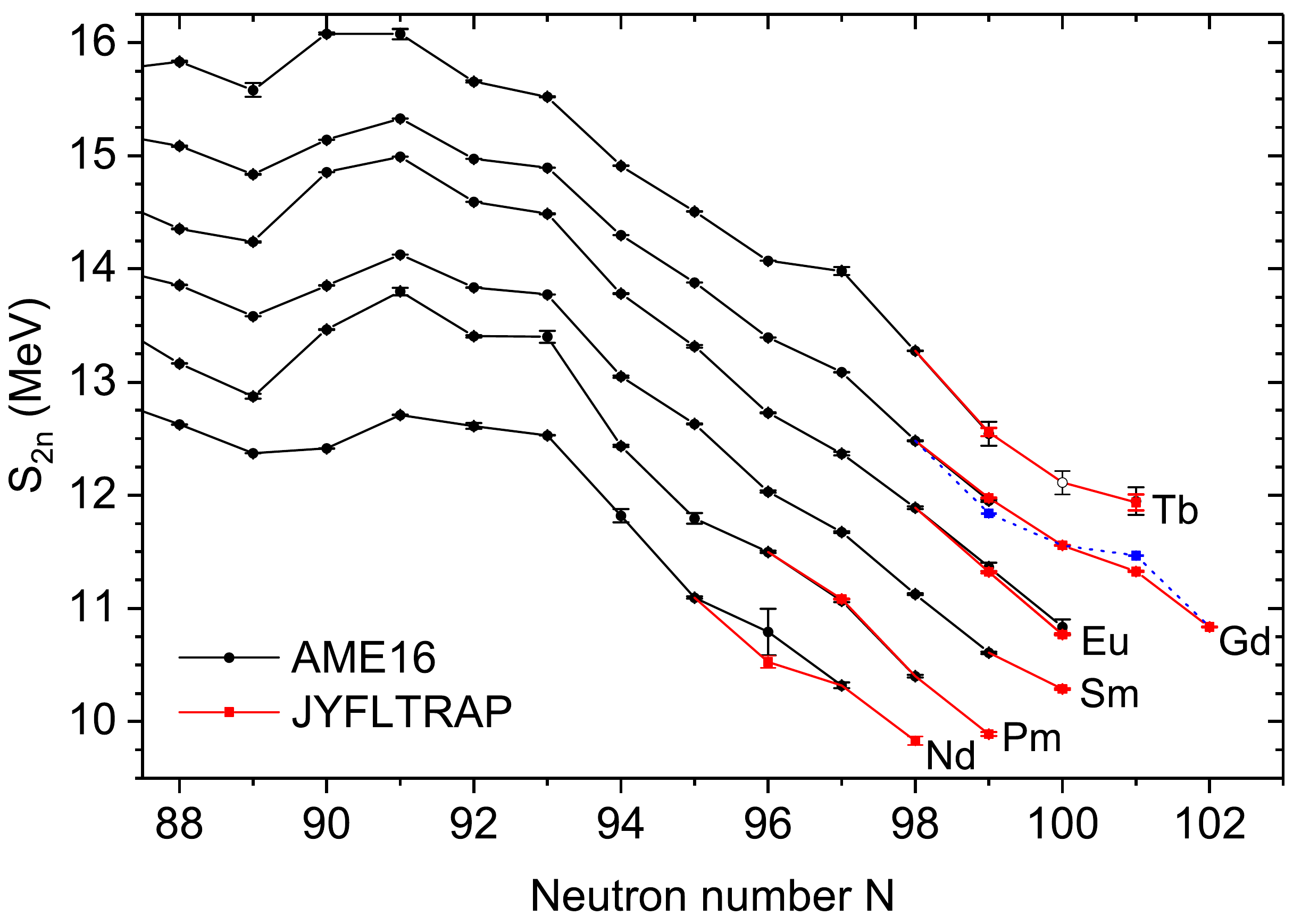}
\caption{(Color online) Two-neutron separation energies $S_{2n}$ from this work (red) together with the experimental (solid black circles) values and an extrapolated value for $^{165}$Tb (open black circle) from AME16 \cite{AME16}. The dashed blue lines indicate the values assuming the ground state of $^{163}$Gd was measured in this work.}
\label{figure_s2n}
\end{figure}

Nuclear structure far from stability can be probed via two-neutron separation energies $S_{2n}$ \cite{LunneyMasses}. They usually exhibit smooth trends except at shell closures or when there is a change in the nuclear structure, such as the onset of strong prolate deformation at around $N=90$ seen as a bump in Fig.~\ref{figure_s2n}. This is also observed as a sharp increase in experimental $E(4^+)/E(2^+)$ ratios reaching $\approx 3.3$ in the region $N=92-102$ compatible with a rigid rotor. The strong deformation is also predicted by theoretical models, e.g. FRDM12 yields a maximum deformation ($\beta_2\approx 0.31$ \cite{Moller2016}) for the Gd isotopes at around $N=101-103$. The new $S_{2n}$ values determined in this work show a change in the slope after $N=100$ for the Gd isotopes ($Z=64$). A similar effect is also observed for Tb at $N=100$ and after $N=96$ for the Nd ($Z=60$) chain. Incidentally, a small local maximum is seen in the $E(2^+)$ energies at $N=100$ for Gd and Dy. However, the two-neutron shell-gap energies for $N=100$ are rather low ($<1$~MeV) down to Gd, and do not support the proposed subshell gap at $N=100$ \cite{Satpathy2003,Satpathy2004,Ghorui2012}.

\begin{figure}
\centering
\includegraphics[trim = 24mm 57mm 35mm 22mm, clip,width=\linewidth]{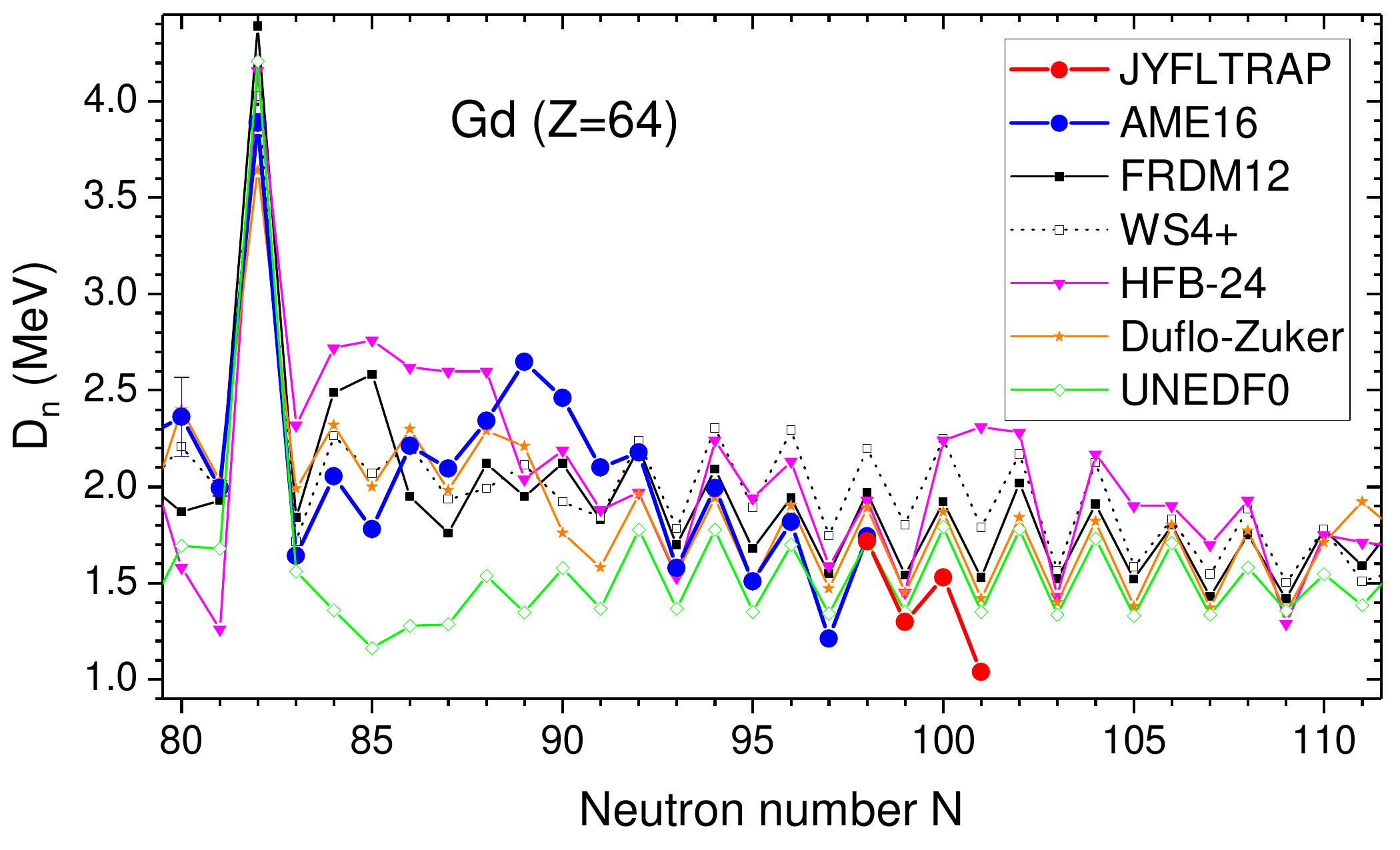}
\caption{(Color online) 
Neutron pairing energies from this work (red circles) and AME16 (blue) in comparison with various theoretical predictions for the Gd isotopes.}
\label{fig:Dn}
\end{figure}

We compared the experimental $S_{2n}$ values to the predictions from various mass models commonly used in $r$-process calculations, such as FRDM12 \cite{Moller2016}, Duflo-Zuker \cite{duflo1995}, and HFB-24 \cite{Goriely2013}. These models predict a rather smooth behavior for the $S_{2n}$ values in the region of interest but overestimate them at $N=99$ and 100 by around 0.3 MeV for the studied isotopic chains. None of them suggest changes in the slope in contrast to those observed in this work. Among the other mass models, WS4+ \cite{Wang2014} yields the smallest root-mean-square (rms) error for the studied isotopic $S_{2n}$ chains, 0.12 MeV. UNEDF0 \cite{Kortelainen2010} results in a similar rms error as HFB-24 and FRDM12, $\approx0.4$ MeV, which is much smaller than for SkM and SLy4 also used in the $r$-process calculations in \cite{Martin2016}.  
To further explore the evolution of nuclear structure, we studied neutron pairing energy metrics $D_n(N)=(-1)^{N+1}[S_n(Z,N+1)-S_n(Z,N)]$ \cite{Brown2013}, which is directly related to the empirical neutron pairing gap $\Delta^3(N)=D_n(N)/2$ \cite{Satula1998} also known as the odd-even staggering parameter. These are very sensitive to changes in the nuclear structure, see e.g.~\cite{Hakala2012}. To highlight such change, Fig.~{\ref{fig:Dn}} shows the impact of our new mass values on $D_n$ for neutron-rich Gd isotopes, an isotopic chain extensively studied \cite{Jones2005, Ghorui2012, wu2017, Yokoyama2017} for its possible change in nuclear structure. Whereas $N$ = 82 presents as a clear peak, nothing is observed at $N$ = 100 to support the existence of a subshell closure. More interestingly, neutron pairing is much weaker than predicted by theoretical models when approaching the midshell. The same can be observed for the other isotopic chains as well: the experimental neutron separation energies are systematically lower at $N=98,100$ and $102$, leading to smaller odd-even staggering than predicted by the theoretical models. While there were already some indications of overestimated even-$N$ $S_n$ values from previous measurements in the Tb, Gd, and Sm chains, these were single cases in their respective chains. The new data presented in this Letter establishes this as a trend, and also extends it to the Pm and Nd chains.

\begin{figure}[h]
\includegraphics[trim = 38mm 2mm 38mm 5mm, clip,width=1.0\linewidth]{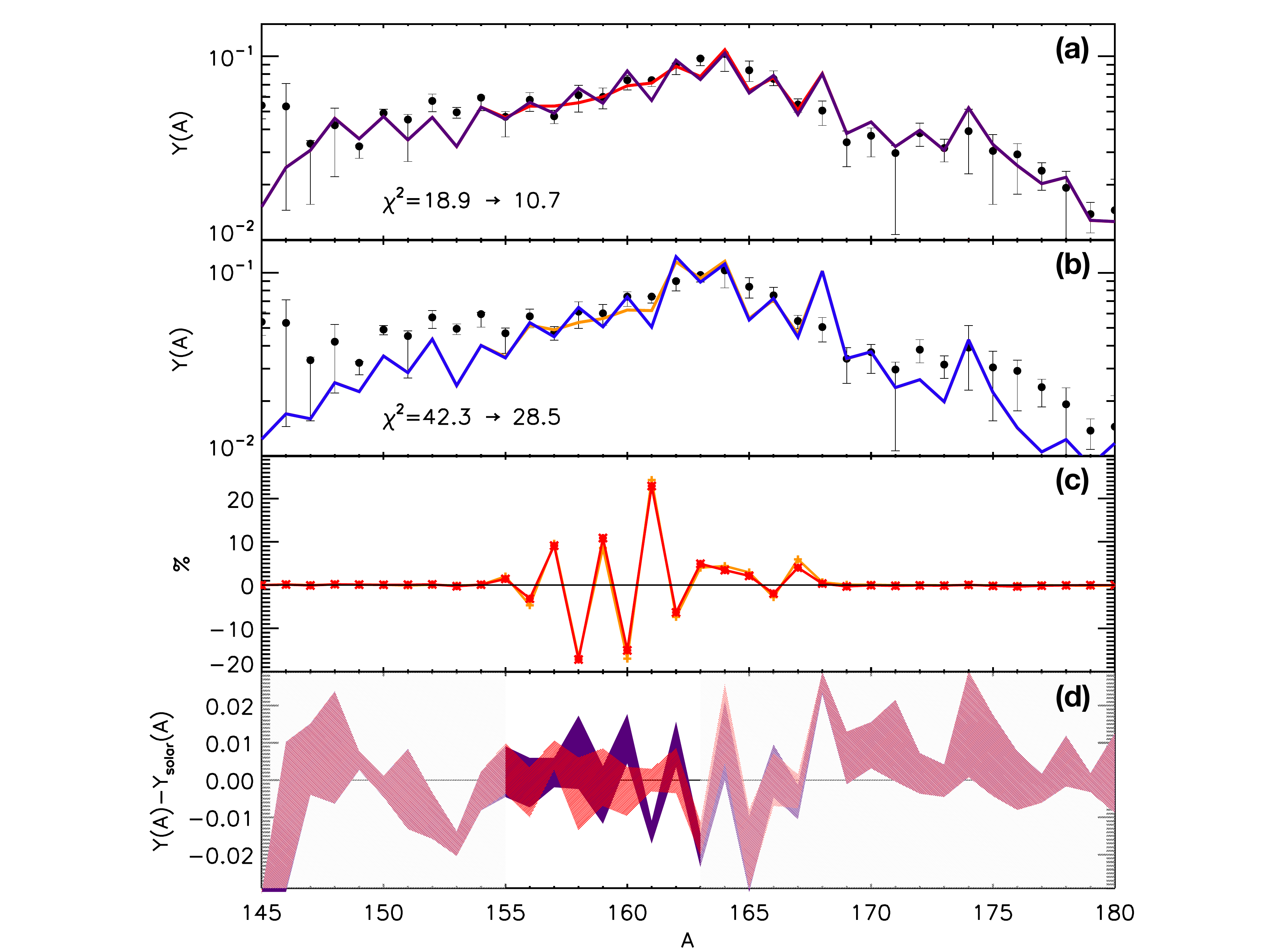}
\caption{(Color online) Solar $r$-process abundances \cite{Arnould2007} (black circles) in comparison with the calculations using the experimental AME16 \cite{AME16} + FRDM12 masses \cite{Moller2016} (blue/purple lines) and the new masses from this work (orange/red lines) for representative trajectories (a) with fission cycling and (b) without fission cycling. (c) Change, in percent, of the abundance pattern as a result of using the masses from this work. (d) Residuals for scenario (a) based on the mass values from this work (red) and the baseline (purple), where the bands represents the solar abundance uncertainties.} 
\label{abundance}
\end{figure}

We studied the impact of the new masses on the $r$-process for astrophysical conditions of a neutron star merger. The $r$-process simulations proceed as in \cite{Mumpower2016}. Masses and relevant $Q$-values not measured in this work were supplemented with experimental data from AME16 or calculated values from FRDM12. For consistency, calculated and experimental masses were not combined in the calculation of a given $S_{n}$-value. Branching ratios and $\beta$-decay half-lives were taken from NUBASE 2016 \cite{nubase16} or \cite{moller2003}. The neutron-capture rates were calculated with the commonly used TALYS code \cite{talys} with the revised mass data set described above.  For fission product distributions we choose a simple asymmetric split \cite{Mumpower2017} so that fission products fall into the $A\sim 130$ region and the rare earth peak forms entirely via the dynamical formation mechanism of \cite{Surman1997,mumpower2012_RE}. The rare earth region of the final abundance patterns for two different types of merger trajectories, corresponding to conditions expected in the dynamical ejecta and accretion disk wind of the merger environment, appear in Fig. \ref{abundance}.

Fig.~\ref{abundance}(a) shows the results from a representative dynamical ejecta trajectory for a 1.35 solar mass neutron star merger from \cite{mendozatemis2015}. The trajectory initially has a very low electron fraction of $Y_{e} = 0.016$ and low entropy per baryon $s/k_{B} \sim 8$, which rises to $s/k_{B} \sim100$ due to nuclear reheating. The timescale is initially around 40 ms, after which a homologous expansion is assumed \cite{mendozatemis2015}. Up to 90\% of the prompt ejected mass may come from these types of reheated, fission-recycling trajectories which all yield very similar abundances with the mass model used, and are therefore largely independent of the specific astrophysical conditions as discussed in \cite{mendozatemis2015}. As shown in Fig.~\ref{abundance}(a,b,d)  better agreement with the solar abundance pattern is obtained including our new mass values ($\chi^{2}=10.7$) than with the AME16 and FRDM12 values used as a baseline ($\chi^{2}=18.9$). Here, $\chi^{2}$ is defined as $\chi^{2}=\sum{[(Y(A)_{solar} - Y(A)_{calc.})/\sigma(Y(A)_{solar})]^{2}}$, where $\sigma(Y(A)_{solar})$ is the uncertainty of the solar abundances \cite{Arnould2007}. The sum is taken over the mass number range ($A = 154 - 168$) affected by the measurements reported in this Letter, and the simulated abundances $Y(A)_{calc.}$ are scaled to solar over the same range. Furthermore, changes of up to 24\% in the calculated abundances resulting in a general smoothing of the profile can be seen, as highlighted in Fig. \ref{abundance}(c).

To examine whether these effects are an artifact of fission recycling, we consider a second type of trajectory that is less neutron-rich and does not undergo fission recycling. We choose a low-entropy, hot wind $r$-process, parametrized as in \cite{2002PhRvL..89w1101M} with values ($s/k_{B}=10$, timescale = 70 ms, $Y_{e}= 0.15$) consistent with those expected for merger accretion disk winds \cite{2015MNRAS.448..541J}. As seen in Figs. \ref{abundance}.(b-c), the influence of the new masses is notably similar to the fission recycling example.

The nuclei studied in this work are populated at late times in the $r$-process, after $(n,\gamma)-(\gamma,n)$ equilibrium has failed. At this stage, the material is decaying back toward stability and the fine details of the final abundance pattern are set through a competition between neutron capture and $\beta$-decay. Although the present work provides more accurate $Q_\beta$ values relevant for the $\beta$-decays, they do not affect the $\beta$-decay rates because the half-lives are already experimentally known. Thus, the visible shifts in the abundance distribution are due entirely to the influence of the new masses on the recalculated neutron-capture rates, which changed by 10-25$\%$. These rates depend on neutron separation energies but also on the choice of the neutron-capture code. Therefore, the calculations done with the TALYS code should be taken as a representative example of the effect of the new mass values on the $r$-process abundances. However, it can be expected that the effect of the revised neutron separation energies would be rather similar even if a different code was used. The reduced neutron pairing observed in this work, i.e. smaller odd-even staggering in the neutron separation energies, is not predicted by FRDM12 or other mass models typically used for the $r$-process calculations (see Fig.~\ref{fig:Dn}). As a result, the final calculated $r$-process abundances are smoother than the baseline calculation done with AME16+FRDM12. More mass measurements are anticipated to test if the see-saw pattern in the abundances at heavier mass numbers is due to the used theoretical mass values.

In this work, we have determined nuclear binding energies for $^{158}\mathrm{Nd}$, $^{160}\mathrm{Pm}$, $^{162}\mathrm{Sm}$, and $^{164-166}\mathrm{Gd}$ for the first time, and improved the precisions for $^{156}\mathrm{Nd}$, $^{158}\mathrm{Pm}$, $^{162,163}\mathrm{Eu}$, and $^{164}\mathrm{Tb}$. Neutron pairing in the very neutron-rich isotopes has been found to be weaker than predicted by the theoretical models commonly used in $r$-process calculations. The data do not support the existence of a subshell closure at  $N=100$. This is in agreement with the conclusions made in Refs.~\cite{wu2017,Yokoyama2017}. While the changes in the slopes of the $S_{2n}$ values coincide with the observed changes in the $E(2^+)$ energies ~\cite{Jones2005,Patel2014}, they may also be due to the approaching maximum deformation in the midshell or reduced neutron pairing. Here, further spectroscopic studies would yield valuable information. The impact of the new mass values on the $r$-process abundance pattern in the rare-earth region has been investigated for two representative neutron star merger trajectories. Changes of up to 24$\%$ and a smoothening of the abundance pattern has been observed for both scenarios. Furthermore, the calculated abundances are now closer to the solar $r$-process abundances. The results of this work highlight the need for accurate mass values in the rare-earth region and provide valuable data to improve theoretical mass models needed for experimentally unreachable nuclei in the $r$ process. This is increasingly important in the era of multi-messenger observations from neutron-star mergers. 

This work has been supported by the Academy of Finland under the Finnish Centre of Excellence Programme 2012-2017 (Nuclear and Accelerator Based Physics Research at JYFL) and by the National Science Foundation (NSF) Grants No. PHY-1419765 and PHY-1713857. A.K., D.N., L.C., and T.E. acknowledge support from the Academy of Finland under projects No. 275389 and 295207 . M.M. carried out this work under the auspices of the National Nuclear Security Administration of the U.S. Department of Energy at Los Alamos National Laboratory under Contract No. DE-AC52-06NA25396. R.S. work is funded in part by the DOE Office of Science under contract DE-SC0013039.


%

\end{document}